%% file: main.tex
\documentclass[conference]{IEEEtran}
\IEEEoverridecommandlockouts

\usepackage{cite}
\usepackage{amsmath,amssymb,amsfonts}
\usepackage{graphicx}
\usepackage{textcomp}
\usepackage{xcolor}
\usepackage{booktabs}
\usepackage{array}
\usepackage{tabularx}
\usepackage[caption=false,font=footnotesize]{subfig}
\usepackage{url}
\usepackage{dblfloatfix}
\usepackage{orcidlink}
\usepackage{tikz}
\usetikzlibrary{positioning, fit, backgrounds, arrows.meta, shadows, calc}

\graphicspath{{./}}

\newcolumntype{Y}{>{\raggedright\arraybackslash}X}
\def\BibTeX{{\rm B\kern-.05em{\sc i\kern-.025em b}\kern-.08em
    T\kern-.1667em\lower.7ex\hbox{E}\kern-.125emX}}

\begin{document}

\title{Remote Sensing-Based Hybrid Statistical–Machine Learning Ensemble Modeling of Yellowfin Tuna Catch in the Western Equatorial Indian Ocean}
\author{
\IEEEauthorblockN{Avishka Wijepala,\orcidlink{0009-0003-6716-6099}}
\IEEEauthorblockA{
\textit{Department of Mathematics}\\
\textit{University of Ruhuna}\\
Matara, Sri Lanka\\
wijepala12479@usci.ruh.ac.lk
}

\and

\IEEEauthorblockN{Sharukshan Niranjan}
\IEEEauthorblockA{\textit{Dept. of Electrical and}\\
\textit{Electronic Eng.}\\
\textit{University of Peradeniya}\\
Peradeniya, Sri Lanka\\
n.sharukshan@gmail.com}

\and

\IEEEauthorblockN{Lisitha Abeysekara}
\IEEEauthorblockA{
\textit{Department of Computer Eng.}\\
\textit{University of Peradeniya}\\
Peradeniya, Sri Lanka\\
lisithaabeysekara@gmail.com
}
\and
\IEEEauthorblockN{Pasindu Weerasinghe}
\IEEEauthorblockA{\textit{Dept. of Electrical and}\\
\textit{Electronic Eng.}\\
\textit{University of Peradeniya}\\
Peradeniya, Sri Lanka\\
e20429@eng.pdn.ac.lk}
}

\maketitle

\begin{abstract}
Yellowfin tuna (\textit{Thunnus albacares}) is an economically important pelagic species in the Indian Ocean, but its monthly catch variability is difficult to predict because reported catch is influenced by ocean productivity, thermal habitat, monsoon seasonality, fishing-ground characteristics, and fleet behavior. This study develops a remote sensing-based statistical--machine learning stacked ensemble for modeling monthly Yellowfin tuna catch across five fishing grounds in the western equatorial Indian Ocean during 2003--2024. Monthly fishery records were integrated with satellite-derived chlorophyll-a concentration (CHL) and sea surface temperature (SST). Linear mixed models (LMMs) and a generalized additive mixed model (GAMM) were used to represent interpretable current and lagged environmental effects, while Random Forest models captured nonlinear interactions among lagged, rolling, seasonal, anomaly, and spatial predictors. Expanding-window validation over the pre-COVID evaluation period from 2008 to 2019 was used to reduce temporal leakage. Among individual models, the GAMM achieved the best log-scale performance with $R^2_{\log}=0.137$ and RMSE$_{\log}=1.284$, while RF-E6 achieved the lowest kilogram-scale RMSE. The non-negative Ridge stacked ensemble improved log-scale performance to $R^2_{\log}=0.177$, RMSE$_{\log}=1.254$, and MAE$_{\log}=1.000$. These results show that satellite-derived CHL and SST contain useful but incomplete predictive information for monthly catch modeling.
\end{abstract}


\section{Introduction}

Yellowfin tuna (\textit{Thunnus albacares}) is a highly migratory pelagic species distributed throughout tropical and subtropical oceans. In the Indian Ocean, it supports offshore fisheries, seafood export chains, and coastal livelihoods. However, catch varies substantially across months and fishing grounds because tuna availability is influenced by thermal habitat, prey fields, productivity gradients, seasonal circulation, and fishing activity. The western equatorial Indian Ocean is especially dynamic because monsoon forcing modifies surface temperature, productivity, and regional circulation.

\begin{figure}[!t]
\centering
\includegraphics[width=0.95\columnwidth,height=0.21\textheight,keepaspectratio]{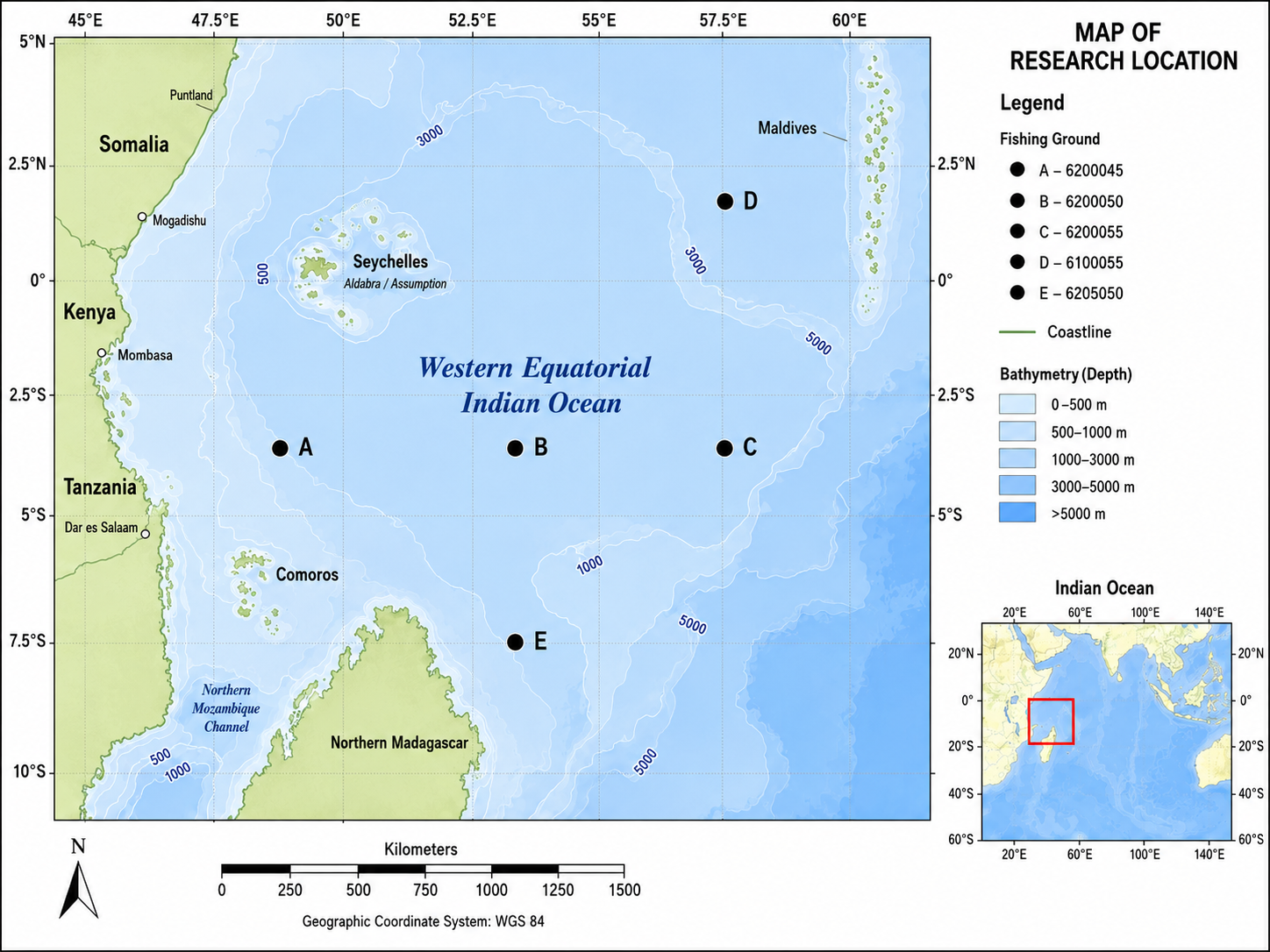}
\caption{Study area and five fishing grounds.}
\label{fig:study_area}
\end{figure}

\begin{figure*}[!t]
\centering
\resizebox{0.92\textwidth}{!}{
    \input{figure.tex}
}
\caption{Stacked-ensemble modeling architecture.}
\label{fig:workflow}
\end{figure*}
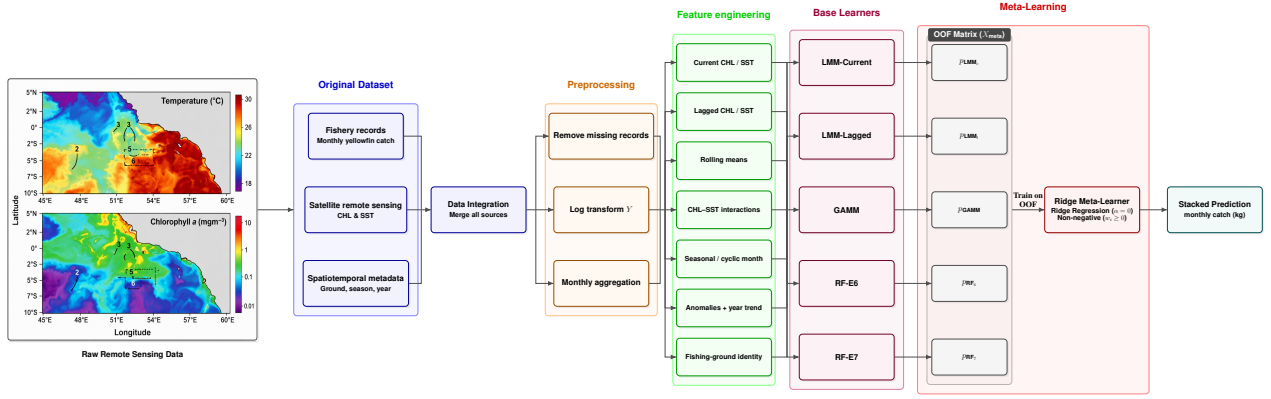

Satellite remote sensing provides consistent, spatially continuous observations of marine environments and is widely used in fisheries monitoring. CHL is typically treated as a proxy for phytoplankton biomass and surface productivity, while SST reflects thermal habitat constraints and large-scale ocean variability. Prior studies have linked tuna catch or CPUE with oceanographic variables such as CHL, SST, currents, sea surface height, mixed-layer depth, fishing depth, and climate indices \cite{maddumage2023effect,wiryawan2020catch, wilson2022satellite,arrizabalaga2015global,lan2013effects,han2022environment}. Machine-learning and environmental covariates have also improved predictive performance when properly validated without temporal leakage \cite{holmes2021improving,giri2025short}. Recent remote-sensing work further suggests that combining spatial and temporal features enhances pattern extraction from satellite data \cite{wijenayake2025precision,wijenayake2026mamba}, supporting the use of multi-temporal feature integration. Accordingly, this study adopts a tabular feature-engineering framework incorporating current, lagged, rolling, seasonal, anomaly, interaction, and fishing-ground predictors derived from CHL and SST to model monthly yellowfin tuna catch.
However, total catch is not a direct proxy for abundance, as it reflects both environmental conditions and fishing effort, including vessel activity, gear configuration, decision-making, market forces, regulations, and reporting practices. CPUE is generally more appropriate for abundance inference \cite{maunder2004standardizing}. Thus, this study focuses on the predictive information contained in satellite-derived variables and spatio-temporal metadata for monthly catch modeling, rather than attempting full abundance reconstruction.
The main contributions are threefold. First, seasonal variability in CHL, SST, and Yellowfin tuna catch is characterized across five fishing grounds. Second, current and lagged environmental effects are evaluated using LMMs and a GAMM. Third, a hybrid statistical--machine learning stacked ensemble is developed to combine interpretable statistical learners with nonlinear machine-learning learners. The full modeling architecture is summarized in Fig.~\ref{fig:workflow}.

\section{Dataset and Experimental Design}

\subsection{Study Data and Satellite Preprocessing}

The dataset consists of monthly yellowfin tuna catch records from 2003 to 2024 across five fishing grounds in the western equatorial Indian Ocean, as shown in Fig.~\ref{fig:study_area}. The response variable was monthly reported catch weight in kilograms. Fishery records were merged with satellite-derived CHL and SST observations using a common ground-month temporal unit.

Monthly CHL was derived from satellite ocean-color chlorophyll-a retrievals, which provide spatially consistent information on upper-ocean biological productivity \cite{mcclain2009decade}. Monthly SST was obtained from the Optimum Interpolation Sea Surface Temperature version 2.1 product \cite{huang2021improvements}. Both variables were temporally aggregated to monthly scale and spatially matched to the fishing-ground locations. Where a fishing ground overlapped multiple satellite pixels, the monthly spatial mean was used to represent the environmental condition of that fishing ground.

The merged panel contained 1,320 ground-month observations, corresponding to five fishing grounds observed over 264 months. Feature sets requiring 12-month lags retained 1,260 observations after removing the first 12 months within each fishing ground. Ground-month records with missing catch, CHL, or SST after merging were excluded from model training and evaluation. Lagged and rolling predictors were computed separately within each fishing ground to prevent cross-ground contamination and temporal leakage.

The catch distribution was strongly right-skewed, ranging from 1.91 kg to 9,502,068 kg, with a median of 431,397.5 kg and a mean of 837,522.8 kg. To reduce the influence of extreme catch months, the response was transformed as
\begin{equation}
Y_{it}=\log(C_{it}+1),
\label{eq:logcatch}
\end{equation}
where $C_{it}$ is the reported catch at fishing ground $i$ during month $t$. CHL ranged from 0.0589 to 0.5022 mg m$^{-3}$, while SST ranged from 24.422$^\circ$C to 31.897$^\circ$C.

\begin{table}[!t]
\caption{Input Variables and Predictor Groups}
\label{tab:variables}
\centering
\scriptsize
\begin{tabularx}{\columnwidth}{@{}p{0.24\columnwidth}p{0.30\columnwidth}Y@{}}
\toprule
Category & Variables & Modeling role \\
\midrule
Response & Catch, $\log(C+1)$ & Stabilized skewed catch variation \\
Remote sensing & CHL, SST & Productivity and thermal habitat \\
Temporal & Month, season, year & Seasonal and inter-annual effects \\
Spatial & Fishing ground & Ground-level differences \\
Lagged predictors & CHL/SST lags & Delayed ecological response \\
Derived predictors & Rolling means,interactions, anomalies & Nonlinear and persistent effects \\
\bottomrule
\end{tabularx}
\end{table}

\subsection{Experimental Design}

The analysis was organized into two connected experiments. The first examined environmental structure and interpretability using descriptive statistics, Spearman correlation analysis, current-month LMMs, lagged LMMs, and a GAMM. This stage evaluated whether lagged satellite variables and nonlinear response functions were important for catch modeling.

The second experiment developed machine-learning and stacked-ensemble models using seven progressively enriched feature sets. Starting with CHL and SST (E1), subsequent experiments added temporal variables, lagged environmental predictors, rolling averages, interaction terms, fishing-ground identity, and long-term trend and anomaly variables, culminating in E7. Among the evaluated models, including Random Forest, XGBoost, support vector regression, and a median baseline, Random Forest provided the strongest machine-learning benchmark.

\section{Methodology}

\subsection{Statistical Base Learners}

Two LMMs were fitted using log-transformed catch as the response. Fishing ground was included as a random intercept to account for persistent spatial differences among grounds\cite{lindstrom1988newton}. The current-month LMM was defined as
\begin{equation}
Y_{it}=\beta_0+\beta_1\mathrm{CHL}_{it}+\beta_2\mathrm{SST}_{it}+\beta_3\mathrm{Season}_t+u_i+\epsilon_{it},
\label{eq:lmmcurrent}
\end{equation}
where $u_i$ is the fishing-ground random intercept and $\epsilon_{it}$ is the residual error. A lagged LMM replaced current CHL and SST with one-month lagged predictors.

The GAMM was specified as
\begin{equation}
Y_{it}=\beta_0+s(\mathrm{CHL}_{i,t-1})+s(\mathrm{SST}_{i,t-1})+\beta_1\mathrm{Season}_t+s(g_i)+\epsilon_{it},
\label{eq:gamm}
\end{equation}
where $s(\cdot)$ denotes a smooth function and $s(g_i)$ is a random-effect smooth for fishing ground. GAMMs are suitable for ecological and fisheries applications because they represent nonlinear environmental responses while retaining interpretability \cite{wood2017generalized}.

\subsection{Machine-Learning Base Learners}

Random Forest was used as the primary machine-learning model because it captures nonlinear effects, interaction structure, threshold behavior, and variable importance without requiring a prespecified response shape. Its use in ecological modeling is well established for modeling complex predictor interactions and ranking variable importance \cite{cutler2007random}. XGBoost and support vector regression were also evaluated as nonlinear alternatives\cite{chen2016xgboost,cortes1995support}. RF-E6 used current and lagged CHL/SST, rolling averages, CHL--SST interactions, seasonal/cyclic temporal encodings, and fishing-ground identity. RF-E7 extended RF-E6 by adding 12-month lags, year trend, and anomaly variables.

The machine-learning models were fitted to the same log-transformed response used in the statistical models. Predictions were back-transformed to kilograms only for practical interpretation and kilogram-scale RMSE calculation.

\subsection{Stacked Meta-Learner}

The final predictive model was a stacked ensemble combining current-month LMM, lagged LMM, GAMM, RF-E6, and RF-E7. Stacking was used because statistical and machine-learning models can make different types of errors. The GAMM captures smooth lagged environmental effects, whereas Random Forest captures nonlinear interactions and threshold structures \cite{wolpert1992stacked}.

A non-negative Ridge meta-learner was trained on out-of-fold predictions. If $\hat{y}_{ij}$ is the out-of-fold prediction from base learner $j$ for observation $i$, the meta-learner estimates coefficients by solving
\begin{equation}
\min_{\beta_0,w_j\geq0}
\sum_i
\left(
y_i-\beta_0-\sum_{j=1}^{J}w_j\hat{y}_{ij}
\right)^2
+\lambda\sum_{j=1}^{J}w_j^2 ,
\label{eq:ridge_objective}
\end{equation}
where $\beta_0$ is the intercept, $w_j$ is the non-negative coefficient, and $\lambda$ is the Ridge penalty parameter. The final stacked prediction is
\begin{equation}
\hat{y}_{i}^{\mathrm{stack}}=\beta_0+\sum_{j=1}^{J}w_j\hat{y}_{ij}, \qquad w_j\geq0.
\label{eq:stacking}
\end{equation}
The coefficients are not required to sum to one because the intercept calibrates the stacked prediction \cite{wolpert1992stacked,friedman2010regularization}.

\subsection{Validation and Evaluation}

All predictive models were evaluated using expanding-window validation over the pre-COVID period (2008–2019), where each model was trained on all available years preceding the test year and evaluated on the subsequent held-out year. This prevents information leakage and approximates an operational forecasting setting. Post-2020 data were excluded from the main comparison due to structural disruptions in reported catch during the COVID-19 period.

Performance was assessed using $R^2_{\log}$, RMSE$_{\log}$, MAE$_{\log}$, and RMSE in kilograms. Log-scale metrics were prioritized because the models were fitted to $\log(C+1)$.

\section{Results and Discussion}

\subsection{Seasonal and Correlation Structure}

Seasonal summaries are reported in Table~\ref{tab:seasonal}. The Southwest Monsoon had the highest mean CHL concentration, 0.2353 mg m$^{-3}$, and the lowest mean SST, 27.014$^\circ$C. Intermonsoon 1 had the lowest mean CHL and the highest mean SST, while Intermonsoon 2 recorded the highest median catch. This mismatch indicates that reported catch is not controlled by simultaneous CHL and SST alone.

\begin{table}[!htbp]
\caption{Seasonal Catch and Environmental Summary}
\label{tab:seasonal}
\centering
\scriptsize
\begin{tabular}{@{}lrrrr@{}}
\toprule
Season & Med. catch & Mean catch & CHL & SST \\
 & (kg) & (kg) & (mg m$^{-3}$) & ($^\circ$C) \\
\midrule
Northeast Monsoon & 409,171.8 & 809,457.4 & 0.1516 & 28.706 \\
Intermonsoon 1 & 394,271.6 & 698,014.8 & 0.1096 & 29.849 \\
Southwest Monsoon & 420,036.0 & 896,651.6 & 0.2353 & 27.014 \\
Intermonsoon 2 & 540,914.4 & 991,566.7 & 0.1480 & 28.682 \\
\bottomrule
\end{tabular}
\end{table}

\begin{figure*}[!t]
\centering
\subfloat[Seasonal SST.]{
\includegraphics[width=0.48\textwidth]{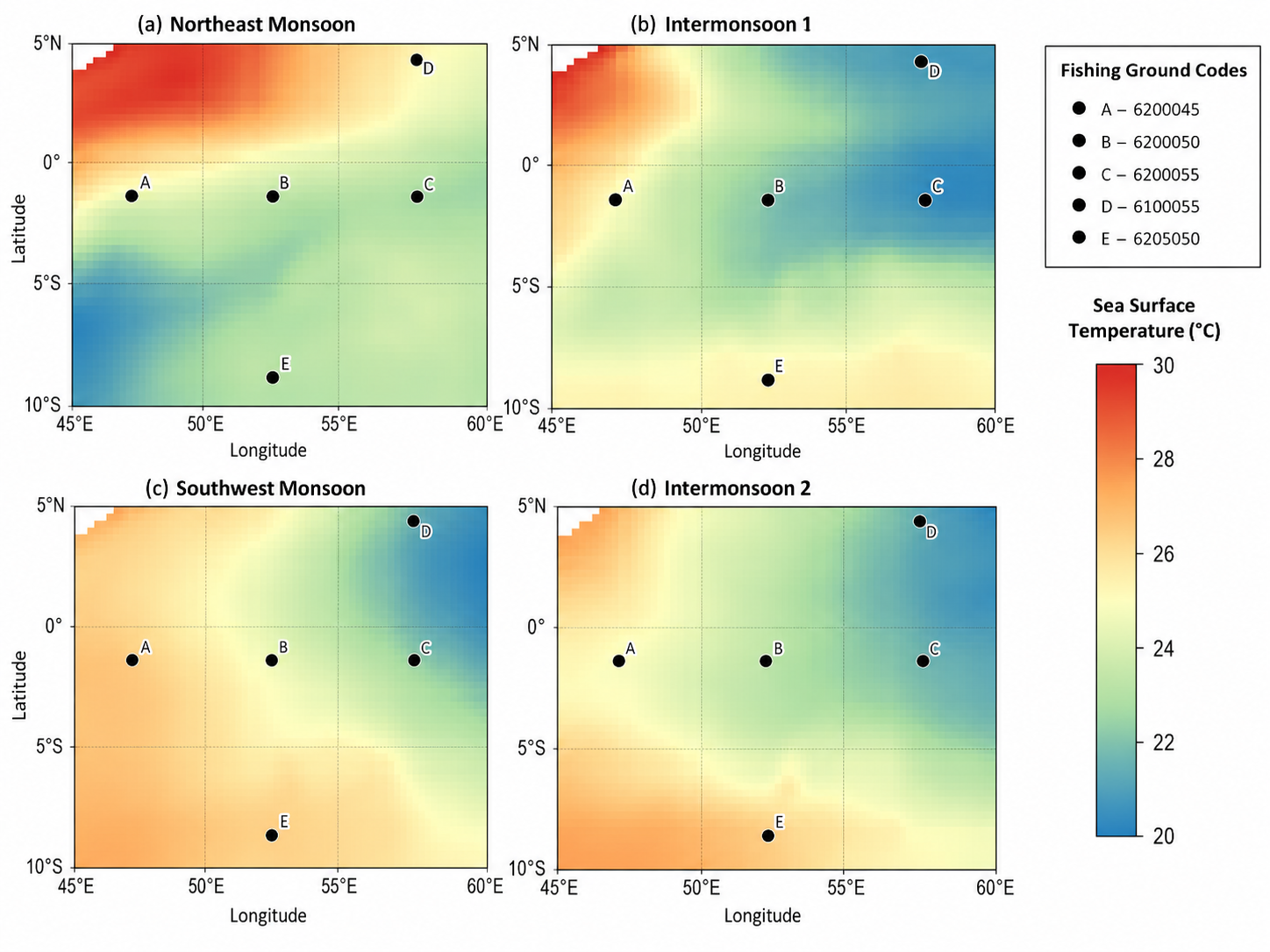}}
\hfill
\subfloat[Seasonal CHL.]{
\includegraphics[width=0.48\textwidth]{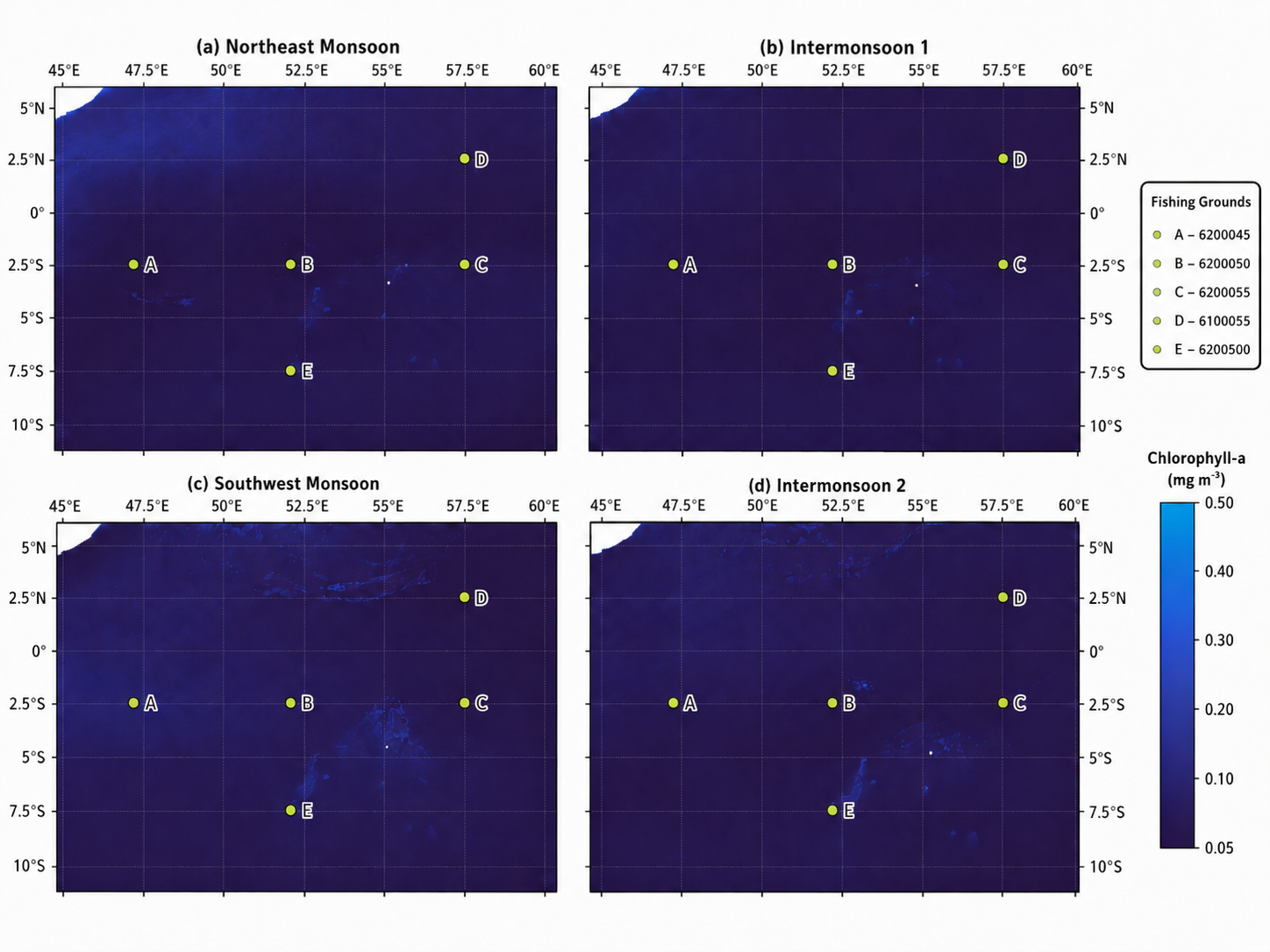}}
\caption{Seasonal SST and CHL conditions.}
\label{fig:env_maps}
\end{figure*}

\begin{figure}[!t]
\centering
\includegraphics[width=0.80\columnwidth]{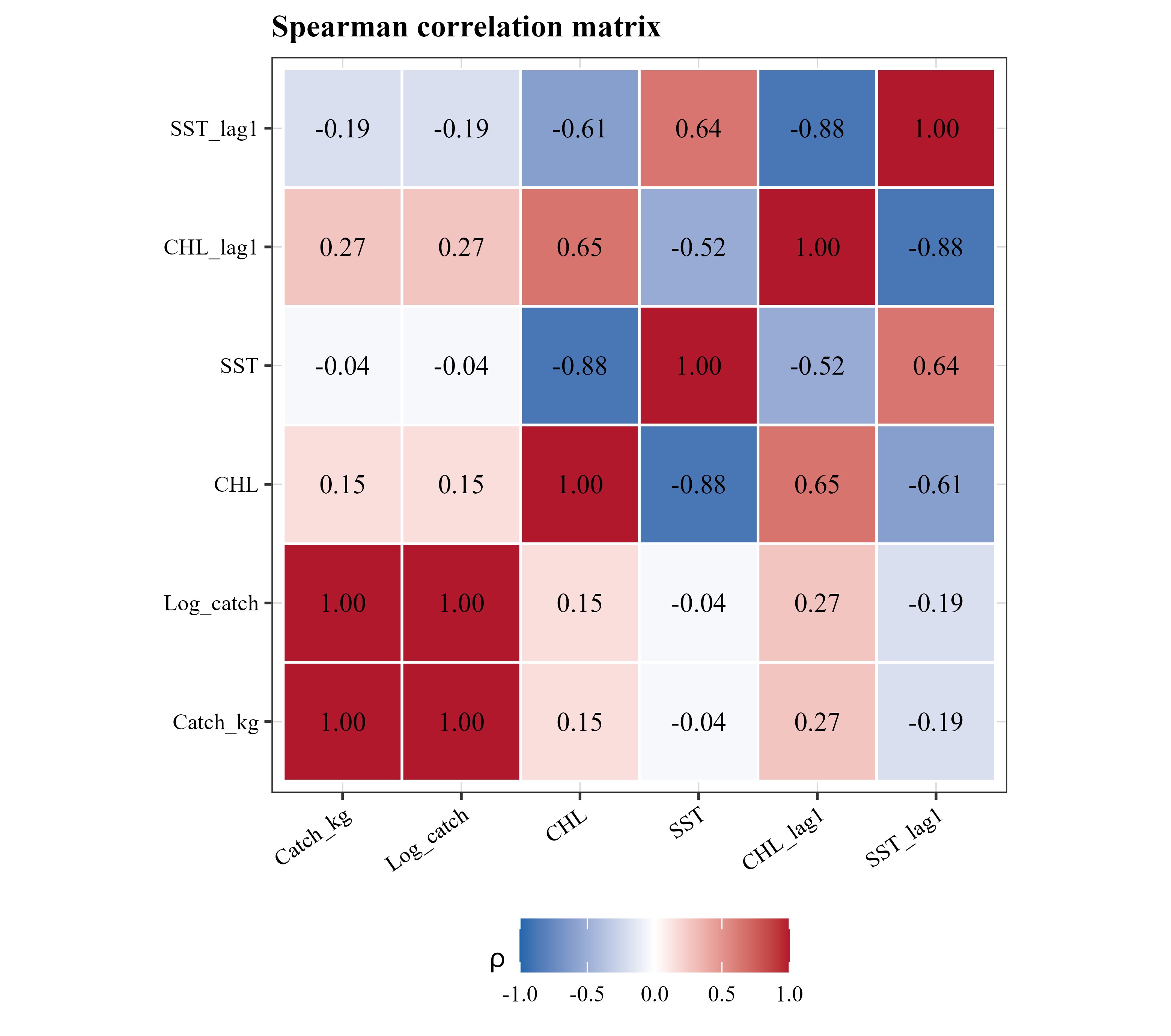}
\caption{Spearman correlation matrix.}
\label{fig:corr}
\end{figure}

The seasonal pattern shows strong monsoon-driven environmental variability, but the catch response is less direct. This supports the interpretation that CHL and SST provide useful environmental signals, while monthly reported catch is also influenced by fishing effort, vessel behavior, and other unobserved operational factors. The Spearman correlation matrix in Fig.~\ref{fig:corr} supports the inclusion of lagged predictors. One-month lagged CHL showed a stronger positive association with catch than current-month CHL.

\subsection{Environmental Effects from the GAMM}

\begin{figure}[!t]
\centering
\includegraphics[width=0.90\columnwidth]{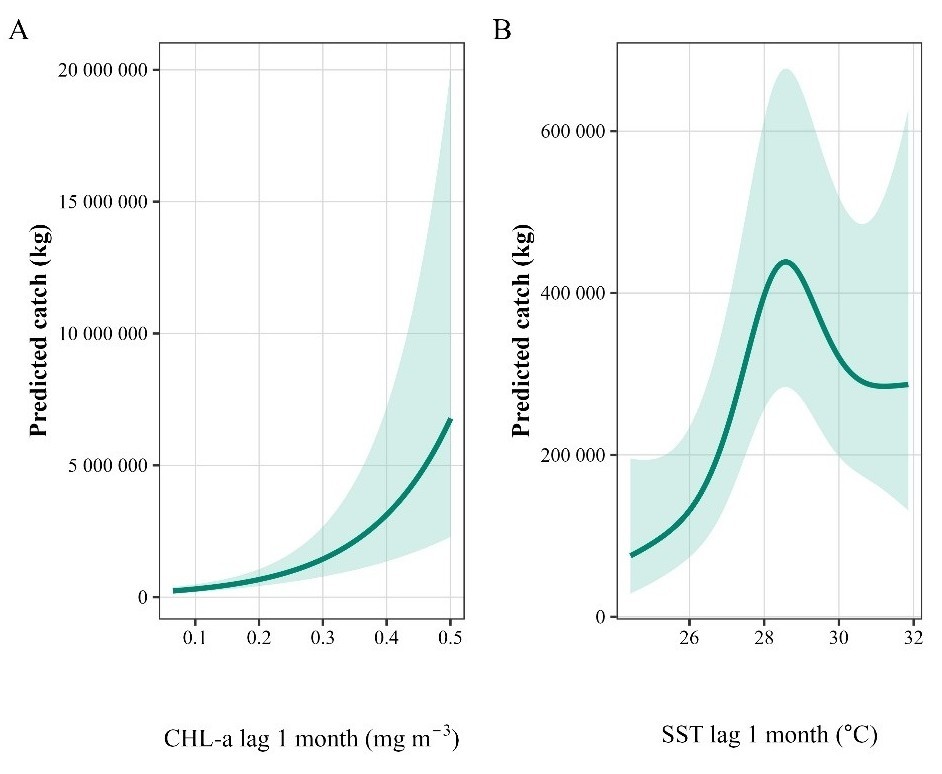}
\caption{GAMM partial effects.}
\label{fig:gamm_effects}
\end{figure}

As shown in Fig.~\ref{fig:gamm_effects}, the final GAMM identified important smooth effects for one-month lagged CHL, one-month lagged SST, and fishing-ground identity. The adjusted $R^2$ was 0.145 and the deviance explained was 15.4\%. These values describe the fitted GAMM structure, while out-of-sample performance is reported in Table~\ref{tab:performance}. The lagged CHL effect was approximately monotonic, suggesting that higher previous-month productivity was associated with higher predicted catch. The lagged SST effect was nonlinear, indicating a favorable thermal range.

These effects are ecologically plausible. CHL may influence tuna availability after a delay because surface productivity must propagate through zooplankton and forage-fish pathways before affecting catchable tuna. Therefore, the positive lagged CHL response in Fig.~\ref{fig:gamm_effects} suggests that previous-month productivity contains useful information for monthly catch prediction. The nonlinear SST response indicates that tuna catch was associated with a favorable thermal range rather than a simple linear temperature effect, which is consistent with the temperature-dependent habitat preference of pelagic tuna. The fishing-ground effect further indicates that spatial identity contains information not fully captured by surface CHL and SST, possibly reflecting persistent differences in fishing access, local circulation, bathymetry, fleet behavior, or ground-specific fishing practices.

\subsection{Predictive Performance and Stacked Ensemble}

Table~\ref{tab:performance} compares the out-of-sample performance of the individual models and the stacked ensemble. Among individual models, the GAMM achieved the strongest log-scale performance, with $R^2_{\log}=0.137$ and RMSE$_{\log}=1.284$. RF-E7 performed better than RF-E6 on log-scale $R^2$, but RF-E6 achieved the lowest kilogram-scale RMSE among all individual models.

\begin{table}[!htbp]
\caption{Out-of-Sample Prediction Performance}
\label{tab:performance}
\centering
\scriptsize
\begin{tabular}{@{}lrrrr@{}}
\toprule
Model & $R^2_{\log}$ & RMSE$_{\log}$ & MAE$_{\log}$ & RMSE (kg) \\
\midrule
Current LMM & 0.132 & 1.288 & 1.022 & 1,246,282 \\
Lagged LMM & 0.108 & 1.305 & 1.042 & 1,211,395 \\
GAMM & 0.137 & 1.284 & 1.029 & 1,207,106 \\
RF-E6 & 0.105 & 1.307 & 1.031 & {1,172,281} \\
RF-E7 & 0.129 & 1.290 & 1.027 & 1,172,571 \\
{Stacked ensemble} & {0.177} & {1.254} & {1.000} & 1,184,824 \\
\bottomrule
\end{tabular}
\end{table}

The stacked ensemble achieved the best log-scale performance, increasing $R^2_{\log}$ from 0.137 to 0.177 (a 29.2\% relative improvement), while also producing the lowest RMSE$_{\log}$ and MAE$_{\log}$. This indicates that combining statistical and machine-learning models improved prediction accuracy on the primary modeling scale. Although RF-E6 achieved the lowest RMSE in kilograms, this is expected because kilogram-scale errors are more sensitive to extreme catches after back-transformation. Thus, log-scale and kilogram-scale metrics provide complementary measures of model performance. The non-negative Ridge meta-learner assigned the largest weights to current LMM (0.281), followed by RF-E7 (0.214), GAMM (0.196), RF-E6 (0.167), and lagged LMM (0.116), indicating that the ensemble benefited from combining complementary models.

\subsection{Interpretation of Modest Predictive Performance}

The modest $R^2_{\log}$ values should be interpreted cautiously, as the response is monthly total catch rather than standardized CPUE. Catch reflects both fish availability and fishing effort, but key effort-related variables (e.g., vessel numbers, fishing days, gear, depth, and FAD use) were unavailable, limiting the separation of environmental and human effects. Hence, the observed performance is scientifically reasonable.

Predictors were restricted to surface CHL, SST, temporal variables, and fishing-ground identity, which capture key productivity and thermal signals but not the full three-dimensional habitat of Yellowfin tuna. Additional oceanographic and fishing-related variables (e.g., SSH anomaly, currents, mixed layer depth, oxygen, salinity, bathymetry, thermocline depth, IOD indices, and vessel activity) could improve performance \cite{druon2017skipjack}. Monthly aggregation may also smooth short-term features and fishing dynamics.

Overall, satellite CHL and SST provide useful but incomplete information for monthly catch prediction. The stacked ensemble improves utilization of this signal but should not be viewed as a full operational tuna abundance forecasting system.

\section{Conclusion}

This study developed a remote sensing based statistical machine learning stacked ensemble for monthly yellowfin tuna catch modeling in the western equatorial Indian Ocean. Satellite-derived CHL and SST showed strong seasonal structure, and the GAMM identified meaningful effects of lagged CHL, nonlinear lagged SST, and fishing ground identity. Under expanding window validation, the stacked ensemble improved out-of-sample log-scale performance from $R^2_{\log}=0.137$ for the best individual model to $R^2_{\log}=0.177$, with RMSE$_{\log}=1.254$ and MAE$_{\log}=1.000$.

The findings suggest that satellite-derived environmental variables contain measurable predictive information for reported monthly catch. Therefore, the proposed framework is best interpreted as a decision support approach for environmental catch modeling rather than a complete abundance forecasting system. Future work should incorporate effort-standardized CPUE, vessel activity, fishing depth, current velocity, sea surface height anomaly, mixed-layer depth, dissolved oxygen, bathymetry, and climate indices such as the Indian Ocean Dipole.

\footnotesize

\nocite{*}
\bibliographystyle{IEEEtran}
\bibliography{references_ieee}
\end{document}

%% file: figure.tex

\begin{tikzpicture}[
    >=Latex,
    base/.style={align=center, minimum height=2.8cm, minimum width=5.8cm, rounded corners=2mm, thick, draw, drop shadow={shadow xshift=3pt, shadow yshift=-3pt, opacity=0.15}, font=\sffamily, inner sep=8pt},
    input/.style={base, fill=blue!5, draw=blue!60!black},
    preprocess/.style={base, fill=orange!5, draw=orange!60!black},
    feature/.style={base, fill=green!5, draw=green!60!black},
    learner/.style={base, fill=purple!5, draw=purple!60!black},
    meta/.style={base, fill=red!5, draw=red!60!black},
    output/.style={base, fill=teal!5, draw=teal!60!black},
    oofcell/.style={align=center, minimum height=2.2cm, minimum width=4.6cm, rounded corners=2mm, thick, draw=gray!60!black, fill=gray!8, drop shadow={shadow xshift=3pt, shadow yshift=-3pt, opacity=0.15}, font=\sffamily, inner sep=6pt},
    groupbox/.style={draw, thick, rounded corners=2mm, inner sep=16pt, fill opacity=0.04},
    input_grp/.style={groupbox, draw=blue!50, fill=blue},
    prep_grp/.style={groupbox, draw=orange!50, fill=orange, inner xsep=6pt},
    feat_grp/.style={groupbox, draw=green!50, fill=green, inner xsep=6pt},
    learn_grp/.style={groupbox, draw=purple!50, fill=purple},
    oof_grp/.style={groupbox, draw=gray!55, fill=gray, inner xsep=8pt},
    meta_grp/.style={groupbox, draw=red!50, fill=red},
    header/.style={font=\bfseries\sffamily\LARGE, align=center},
    ooftab/.style={rounded corners=1.2mm, fill=gray!55!black, text=white, font=\bfseries\sffamily\Large, inner xsep=10pt, inner ysep=5pt, anchor=center},
    arrow/.style={-{Latex[scale=1.4]}, thick, draw=black!70},
    line/.style={thick, draw=black!70}
]

    \node[draw=black!50, thick, rounded corners=1mm, inner sep=0pt, anchor=east, drop shadow={shadow xshift=3pt, shadow yshift=-3pt, opacity=0.15}] (map) at (-6, 0) {\includegraphics[height=16cm]{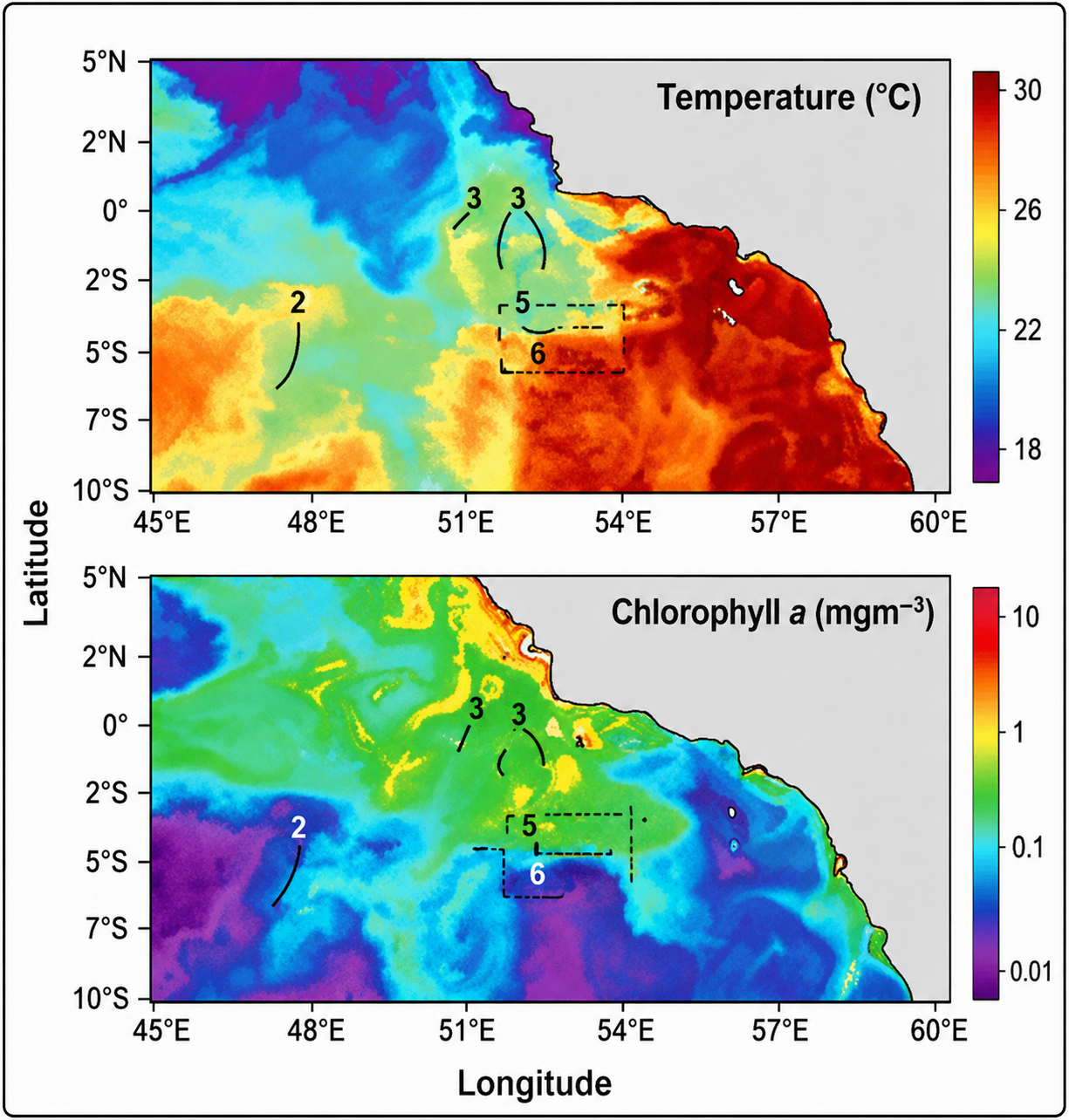}};
    \node[font=\Large\bfseries\sffamily, below=0.5cm of map] {Raw Remote Sensing Data};

    \node[input] (fish) at (0, 4.5)  {\Large\textbf{Fishery records}\\[0.5em] \large\textbf{Monthly yellowfin catch}};
    \node[input] (sat)  at (0, 0)    {\Large\textbf{Satellite remote sensing}\\[0.5em] \large\textbf{CHL \& SST}};
    \node[input] (meta) at (0, -4.5) {\Large\textbf{Spatiotemporal metadata}\\[0.5em] \large\textbf{Ground, season, year}};

    \node[input] (integ) at (7.5, 0) {\Large\textbf{Data Integration}\\[0.5em] \large\textbf{Merge all sources}};

    \node[preprocess] (prep1) at (15.0, 4.5)  {\Large\textbf{Remove missing records}};
    \node[preprocess] (prep2) at (15.0, 0)    {\Large\textbf{Log transform $Y$}};
    \node[preprocess] (prep3) at (15.0, -4.5) {\Large\textbf{Monthly aggregation}};

    \node[feature, minimum height=2.3cm] (f1) at (22.5,  9.0) {\large\textbf{Current CHL / SST}};
    \node[feature, minimum height=2.3cm] (f2) at (22.5,  6.0) {\large\textbf{Lagged CHL / SST}};
    \node[feature, minimum height=2.3cm] (f3) at (22.5,  3.0) {\large\textbf{Rolling means}};
    \node[feature, minimum height=2.3cm] (f4) at (22.5,  0)   {\large\textbf{CHL--SST interactions}};
    \node[feature, minimum height=2.3cm] (f5) at (22.5, -3.0) {\large\textbf{Seasonal / cyclic month}};
    \node[feature, minimum height=2.3cm] (f6) at (22.5, -6.0) {\large\textbf{Anomalies + year trend}};
    \node[feature, minimum height=2.3cm] (f7) at (22.5, -9.0) {\large\textbf{Fishing-ground identity}};

    \node[learner] (lmm_curr) at (30.0, 9.0)  {\Large\textbf{LMM-Current}};
    \node[learner] (lmm_lag)  at (30.0, 4.5)  {\Large\textbf{LMM-Lagged}};
    \node[learner] (gamm)     at (30.0, 0)    {\Large\textbf{GAMM}};
    \node[learner] (rf6)      at (30.0, -4.5) {\Large\textbf{RF-E6}};
    \node[learner] (rf7)      at (30.0, -9.0) {\Large\textbf{RF-E7}};

    \node[oofcell] (o1) at (37.5,  9.0) {\Large\textbf{$p_{\text{LMM}_c}$}};
    \node[oofcell] (o2) at (37.5,  4.5) {\Large\textbf{$p_{\text{LMM}_l}$}};
    \node[oofcell] (o3) at (37.5,  0)   {\Large\textbf{$p_{\text{GAMM}}$}};
    \node[oofcell] (o4) at (37.5, -4.5) {\Large\textbf{$p_{\text{RF}_6}$}};
    \node[oofcell] (o5) at (37.5, -9.0) {\Large\textbf{$p_{\text{RF}_7}$}};

    \node[meta] (ridge) at (45.0, 0) {\Large\textbf{Ridge Meta-Learner}\\[0.5em] \large\textbf{Ridge Regression ($\alpha = 0$)}\\\large\textbf{Non-negative ($w_i \ge 0$)}};

    \node[output] (stacked) at (52.5, 0) {\Large\textbf{Stacked Prediction}\\[0.5em] \large\textbf{monthly catch (kg)}};

    \begin{scope}[on background layer]
        \node[input_grp, fit=(fish) (sat) (meta)] (box_in) {};
        \node[header, above=0.7cm of box_in, text=blue!80!black] {Original Dataset};

        \node[prep_grp, fit=(prep1) (prep2) (prep3)] (box_prep) {};
        \node[header, above=0.7cm of box_prep, text=orange!80!black] {Preprocessing};

        \node[feat_grp, fit=(f1) (f7)] (box_feat) {};
        \node[header, above=0.7cm of box_feat, text=green!80!black] {Feature engineering};

        \node[learn_grp, fit=(lmm_curr) (lmm_lag) (gamm) (rf6) (rf7)] (box_learn) {};
        \node[header, above=0.7cm of box_learn, text=purple!80!black] {Base Learners};

        \node[oof_grp, fit=(o1) (o5)] (box_oof) {};
        \node[meta_grp, fit=(box_oof) (ridge)] (box_meta) {};
        \node[header, above=0.7cm of box_meta, text=red!80!black] {Meta-Learning};
    \end{scope}

    \node[ooftab] at (box_oof.north) {OOF Matrix ($X_{\text{meta}}$)};

    \draw[arrow] (map.east) -- (box_in.west);
    \coordinate (merge_in) at (4.0, 0);
    \draw[line] (fish.east) -| (merge_in);
    \draw[line] (meta.east) -| (merge_in);
    \draw[line] (sat.east) -- (merge_in);
    \draw[arrow] (merge_in) -- (integ.west);
    \coordinate (split_prep) at (11.0, 0);
    \draw[line] (integ.east) -- (split_prep);
    \draw[arrow] (split_prep) |- (prep1.west);
    \draw[arrow] (split_prep) -- (prep2.west);
    \draw[arrow] (split_prep) |- (prep3.west);
    \coordinate (merge_prep_out) at (18.6, 0);
    \draw[line] (prep1.east) -| (merge_prep_out);
    \draw[line] (prep3.east) -| (merge_prep_out);
    \draw[line] (prep2.east) -- (merge_prep_out);
    \coordinate (split_feat_in) at (18.9, 0);
    \draw[line] (merge_prep_out) -- (split_feat_in);
    \draw[arrow] (split_feat_in) |- (f1.west);
    \draw[arrow] (split_feat_in) |- (f2.west);
    \draw[arrow] (split_feat_in) |- (f3.west);
    \draw[arrow] (split_feat_in) -- (f4.west);
    \draw[arrow] (split_feat_in) |- (f5.west);
    \draw[arrow] (split_feat_in) |- (f6.west);
    \draw[arrow] (split_feat_in) |- (f7.west);

    \coordinate (fbus) at (26.35, 0);
    \draw[line] (f1.east) -| (fbus);
    \draw[line] (f2.east) -| (fbus);
    \draw[line] (f3.east) -| (fbus);
    \draw[line] (f4.east) -- (fbus);
    \draw[line] (f5.east) -| (fbus);
    \draw[line] (f6.east) -| (fbus);
    \draw[line] (f7.east) -| (fbus);
    \draw[arrow] (fbus) |- (lmm_curr.west);
    \draw[arrow] (fbus) |- (lmm_lag.west);
    \draw[arrow] (fbus) -- (gamm.west);
    \draw[arrow] (fbus) |- (rf6.west);
    \draw[arrow] (fbus) |- (rf7.west);

    \draw[arrow] (lmm_curr.east) -- (o1.west);
    \draw[arrow] (lmm_lag.east)  -- (o2.west);
    \draw[arrow] (gamm.east)     -- (o3.west);
    \draw[arrow] (rf6.east)      -- (o4.west);
    \draw[arrow] (rf7.east)      -- (o5.west);

    \draw[arrow] (box_oof.east) -- (ridge.west) node[midway, above, font=\Large\bfseries, align=center] {Train on\\OOF};
    \draw[arrow] (ridge.east) -- (stacked.west);
\end{tikzpicture}